# Goldstone-like phonon modes in a (111)-strained perovskite

A. Marthinsen[1], S. M. Griffin[2,3], M. Moreau[4], T. Grande[1], T. Tybell[4], and S. M. Selbach[1,*]

[1] Department of Materials Science and Engineering, NTNU – Norwegian University of Science and Technology, 7491 Trondheim, Norway
[2] Molecular Foundry, Lawrence Berkeley National Laboratory, Berkeley, California, USA.
[3] Department of Physics, UC Berkeley, Berkeley, California, USA.
[4] Department of Electronic Systems, NTNU – Norwegian University of Science and Technology, 7491 Trondheim, Norway

**Goldstone modes are massless particles resulting from spontaneous symmetry breaking. Although such modes are found in elementary particle physics as well as in condensed matter systems like superfluid helium, superconductors and magnons - structural Goldstone modes are rare. Epitaxial strain in thin films can induce structures and properties not accessible in bulk and has been intensively studied for (001)-oriented perovskite oxides. Here we predict Goldstone-like phonon modes in (111)-strained SrMnO$_3$ by first-principles calculations. Under compressive strain the coupling between two in-plane rotational instabilities give rise to a Mexican hat shaped energy surface characteristic of a Goldstone mode. Conversely, large tensile strain induces in-plane polar instabilities with no directional preference, giving rise to a continuous polar ground state. Such phonon modes with *U*(1) symmetry could emulate structural condensed matter Higgs modes. The mass of this Higgs boson, given by the shape of the Mexican hat energy surface, can be tuned by strain through proper choice of substrate.**

## I. INTRODUCTION

Spontaneous symmetry breaking (SSB) spans the entire energy landscape of physics with manifestations ranging from high-energy particle collisions to low-temperature phase transitions in condensed matter [1]. When a continuous symmetry is spontaneously broken there exists a massless particle corresponding to a zero frequency collective mode with continuous $U(1)$ symmetry, the so-called `Mexican hat' energy potential [2–4]. This is described by the Goldstone theorem which has been studied in systems ranging from magnons [2] and liquid crystals [5] to Heisenberg ferromagnets [6] and incommensurate phases [7].

Another consequence of SSB is the formation of topological defects, which occur when the symmetry-breaking phase transition also results in a non-trivial change in the topology of the order parameter describing the phase transition. Depending on this topology change and the dimension of the system under consideration, these topological defects can be 2D domain walls, 1D vortices or 'strings' and 0D monopoles [8]. Originally formulated to describe symmetry-breaking phase transitions in the early universe [9,10], several systems in condensed matter have been found to host topological defects such as superfluid $^4$He [11], high temperature superconductors [12,13] and multiferroic h-RMnO$_3$ hexagonal manganites, with defect density predicted by the Kibble-Zurek mechanism [10,11].

Despite their abundance in other phases of matter, few examples of Goldstone modes in crystalline solids have been reported, e.g. artificial layered Ruddlesdon-Popper PbSr$_2$Ti$_2$O$_7$ [14], superconducting Cd$_2$Re$_2$O$_7$ [15,16] and 2D-antiferromagnetic Ca$_2$RuO$_4$ [17]. Additionally, the h-RMnO$_3$ possess a disordered high temperature phase with emergent $U(1)$ symmetry, characteristic for Goldstone modes [18–20].

Coherent epitaxial strain is a powerful way to enhance, control and even induce new functionality, especially in perovskite oxides [21]. Well-known examples include enhanced polarization in BaTiO$_3$ [22], ferroelectricity in SrTiO$_3$ [23] and coexistence of metallic and insulating phases in perovskite manganites [24], as well as multiferroicity in EuTiO$_3$ [25] and SrMnO$_3$ [26]. While most perovskite thin films have been grown along the [001] direction, recent progress has enabled the exploration of a variety of symmetry constraints at oxide interfaces [27]. In particular, the symmetry and interactions across epitaxial (111)-interfaces has led to novel phenomena like exchange bias in $LaNiO_3$-$LaMnO_3$ superlattices [28] and two-dimensional topological insulators [29].

Different crystallographic orientations can give rise to diverging interfacial coupling and resulting properties. Layer stacking sequences in the [111] and [001] directions in perovskites are -[B-$AO_3$]- and –[AO-$BO_2$]-, respectively, giving different interfacial coupling across epitaxial interfaces [30]. Strain in the (111) plane along octahedral faces yields a trigonal distortion of the $BO_6$ octahedra, with tensile and compressive strain inducing out-of-plane compression or elongation, respectively. In contrast, (001) strain parallel or perpendicular to $BO$ bonds in yields a tetragonal distortion, with different crystal field splitting and electronic structure compared to (111) strain [31].

Latent multiferroic $SrMnO_3$ lies at the stability edge between perovskite and 4H hexagonal polytype structure [32], and is hence susceptible to small perturbations which can drive phase transitions [32–36]. Bulk $SrMnO_3$ is a non-polar G-type antiferromagnet [32], while (001) strain has been predicted [37] to induce both ferromagnetism and ferroelectricity, and the latter has been experimentally demonstrated [26]. While dipoles in prototypical ferroelectrics like $BaTiO_3$ are stabilized by covalency between O $2p$ states and the formally empty $d$ states of a $d^0$ B cation [38], the $t_{2g}$ electrons of $Mn^{4+}$ hinder such charge transfer in bulk $SrMnO_3$. However, by reducing electronic repulsions through lattice expansion, e.g. by strain or replacing $Sr^{2+}$ with larger $Ba^{2+}$, charge transfer to the empty $e_g$-states can occur, inducing multiferroicity by "directional $d^0$-ness" [39].

Here we predict both polar and non-polar Goldstone-like modes in (111)-strained $SrMnO_3$ by first principles electronic structure and lattice dynamics calculations, and discuss the necessary structural and chemical factors stabilizing the Goldstone modes. The strain induced SSB phase transitions can potentially host topological vortices. While excitations along the brim of the Mexican hat potential represent Goldstone modes, excitations from the brim to the top of the hat correspond to Higgs modes, and we address the possibility of controlling the Higgs excitation mass by epitaxial strain.

## II. LATTICE INSTABILITIES AND NOTATION

We first address the effect of epitaxial strain on structural instabilities in the high symmetry structures ($R\bar{3}m$ for (111) strain, $P4/mmm$ for (001) strain) using density functional theory calculations. For octahedral rotations we let $\alpha$, $\beta$ and $\gamma$ (Fig. 1(a)) denote rotation about the pseudocubic axes, *x*, *y* and *z*, respectively, while $\alpha'$, $\beta'$ and $\gamma'$ denote rotation about the orthogonal axes along the pseudocubic vectors [$1\bar{1}0$], [$11\bar{2}$] and [111], respectively (Fig. 1(b)), where the two former lie in the (111) plane. Out-of-phase rotational modes are favoured over in-phase modes for all considered strain values, as shown in Fig. 1(c). The degeneracy of the rotational modes is lifted under (111) strain, and the out-of-phase modes split into one rotational mode about the [111] out-of-plane axis and two rotational modes about the [$11\bar{2}$] and [$1\bar{1}0$] in-plane axes [31]. Although the [$11\bar{2}$] direction, pointing towards an octahedral face, and the [$1\bar{1}0$] direction, pointing towards an octahedral edge, are not symmetry equivalent, the rotational modes about these axes are degenerate under all considered strain values. We find that compression favours in-plane rotations over out-of-plane, which are stabilized at about -1% strain, while tensile strain favours out-of-plane rotations. To illustrate the different structural response to (111) strain compared with (001), we show the evolution of rotational and polar phonon modes under (001) strain [40] (Fig. 1d). The splitting of rotational out-of-phase modes into one out-of-plane mode and two degenerate in-plane is analogous to (111) strain, but with the opposite response. Suppression of out-of-plane polarization under (111) compressive strain is geometrically driven since the three oxygens in the $AO_3$ layers are pushed together, hindering Mn displacement [41,42]. Under compressive (001) strain the Poisson elongation of the out-of-plane octahedral axis allows the displacement of Mn.

## III EPITAXIAL STRAIN INDUCED GOLDSTONE-LIKE MODES

### A. Compressive strain

We now consider the energy landscape of the two in-plane rotational modes ($\alpha'$ and $\beta'$) (Fig. 2(a)) under compressive (111) strain, showing the energy lowering from $R\bar{3}m$ as a function of mode amplitude in Fig. 2(b). Both the energy minimizing mode amplitude and the energy lowering increases with increasing compressive (111) strain, implying that dodecahedra are less compressible than octahedra under (111) strain. Condensation of the $\alpha'$ and $\beta'$ modes is equally energy lowering within the energy resolution of our calculations. Under 2% and 4% compressive strain the rotational modes are stabilized by 6 and 12 meV/f.u with respect to the $R\bar{3}m$ phase, respectively.

Due to the six-fold inversion axis ($S_6$), there are six symmetry equivalent crystallographic directions within the $\langle 11\bar{2} \rangle$ and the $\langle 1\bar{1}0 \rangle$ family. Since octahedral rotation about either of these axes are equally energy lowering, we get a total of 12 degenerate energy minima. The coupling between the $\alpha'$ and $\beta'$ rotational modes is very weak, yielding an energy surface with a 'Mexican hat' shape, where the ground state depends only on the distortion amplitude $\propto \alpha^2 + \beta^2 + \gamma^2$, and not the direction of the in-plane rotational axis, as shown in Fig. 2(d). In contrast, under (001) strain the coupling between the in-plane rotational modes $\alpha_{100}$ and $\beta_{010}$ gives a distinct energy minimum along the [110] direction (see Fig. S1).

A continuous degeneracy along the brim of a Mexican hat potential with $U(1)$ symmetry further implies a continuous set of degenerate structures described as a Goldstone-like mode with zero energy barrier to undergo a transition to a new phase with a different in-plane rotational axis (Fig. 2(e)). Importantly, the rotation mode amplitude *increases* with increasing (111) compressive strain for SrMnO$_3$, unlike LaAlO$_3$ where the amplitude decreases due to the different polyhedral compressibilities [31]. Compressive strain can thus serve as a control parameter to tune the rotational mode amplitude in SrMnO$_3$ and consequently the shape of the Mexican hat potential, as discussed further below. The energy difference between the $\alpha'$ and $\beta'$ rotational modes as function of mode amplitude in Fig. 2(c) shows that for increasing amplitudes a preference for β′ develops. This implies that under sufficiently large mode amplitude the degeneracy of the in-plane rotational axes is lifted, giving six energy minima in the brim of the Mexican hat energy potential. Therefore, a further symmetry-breaking phase transition from a $U(1)$ to a $Z_6$ state will occur with increasing mode amplitude upon increasing strain, analogous to the ferroelectric transition in h-RMnO$_3$, which display the same emergent $U(1)$ symmetry [43]. We note that the h-RMnO$_3$ lie on the opposite stability edge with respect to the Goldschmidt tolerance factor compared to SrMnO$_3$.

**B. Tensile strain**

We now turn to *tensile* strain and find that the in-plane rotational modes $α′$ and $β′$ are also imaginary, but suppressed by the out-of-plane $γ′$ mode. However, tensile strain also induces two orthogonal in-plane ferroelectric instabilities where Mn is displaced along the $[1\bar{1}0]$ and $[11\bar{2}]$ axes respectively (Fig. 1(c)). These polar modes become unstable at ~2.5% strain, but competition with the rotational $γ′$ mode suppress them up to 4% strain. The out-of-plane polar mode along [111] is insensitive to (111) strain, remaining stable under both compressive and tensile strain. In contrast, (001) strain softens ferroelectric modes both under compression, **P** || [001], as well as tension, **P** || [100] and **P** || [010] [40] .

We compare the energy lowering of the two orthogonal in-plane polar modes and the competing $γ′$ mode (shown in Fig. 3(c)) from the high symmetry $R\bar{3}m$ phase for tensile (111) strain ranging from 4% to 6% in Fig. 3(a). The two polar modes, each having six symmetry equivalent crystallographic directions, reduce the crystal symmetry from $R\bar{3}m$ to $C2$ (**P**[1$\bar{1}$0]) and $Cm$ (**P**[11$\bar{2}$]) respectively, when disregarding additional rotation. Analogous to the behavior of the in-plane rotational modes under compressive strain, the two polar directions are equally energy lowering with minima differering by ~0.1 meV/f.u. at 4% and ~0.3 meV at 6% strain. However, the energy lowering from the in-plane polar modes is weaker than for the $γ′$ rotational mode, even though the imaginary frequency is lower (Fig.

1(c)). The effect of the rotational mode on the energy landscape along the two polar modes is assessed by freezing in the lowest-energy rotational amplitude and recalculating the energy landscape of the polar distortions (Fig. 3(b)). Condensation of the $\gamma'$ mode does not lift the degeneracy of the two orthogonal polar modes, as there is no out-of-plane displacement of oxygen under $\gamma'$ rotation. However, the $\gamma'$ mode suppresses the energy lowering and amplitude of the orthogonal polar modes. Under 5% and 6% tensile strain, the polar modes retain a finite amplitude under competition with the $\gamma'$ rotational mode, while under 4% strain they are suppressed.

The energy difference between polar off-centering along the [11$\bar{2}$] and the [1$\bar{1}$0] direction increases with off-centering amplitude, progressively favouring displacements along [1$\bar{1}$0], as shown in Fig. 3(d). When the rotational $\gamma'$ mode is included the energy difference between polar off-centering in the two respective directions diminishes due to the reduced symmetry. In the presence of out-of-phase rotations, Mn offsets in consecutive oxygen octahedra are no longer equivalent as they displace towards an octahedral face or edge, hence out-of-phase rotations favor in-plane isotropy between the two polar directions.

The coupling of the two polar modes is visualized for 6% strain in Fig. 3(e-f), in the absence of rotations (3e), and in the presence of rotations frozen with their lowest energy amplitude (3f). As also observed for in-plane rotational modes under compressive strain, there is no directional preference for in-plane polarization under (111) tensile strain, with less than 0.5meV/f.u. variation along the brim of the Mexican hat potential, resulting in polar Goldstone-like modes.

### III. MICROSCOPIC ORIGINS OF GOLDSTONE-LIKE MODES

We now discuss the origin of both the *non-polar* Goldstone-like rotational mode under compression as well as the *polar* Goldstone-like modes under tensile strain. For the *non-polar* Goldstone like modes, decomposition of the total energy into a band energy contribution and electrostatic energy contribution (Fig. S3) show that the α′ and β′ mode have almost identical impact on the two energy contributions for all compressive strains. This is both reflective of the coherent (111) strain behavior, where all Mn-O bonds are equally elastically stressed in absence of rotations, as well as structural changes for any given in-plane rotational axis: as no Mn-O bond is directed within the (111) plane, *any* choice of in-plane rotational will affect *all* Mn-O bonds to some degree. At moderate mode amplitudes, differences between in-plane rotational axes are thereby small, giving rise to a Goldstone-like non-polar phonon mode. Additionally, the band energy, which is increasingly lowered by larger rotations, is dominating the total energy, accounting for the increased stabilization of rotations with compressive strain.

The *polar* Goldstone-like mode are stabilized by both structural and electronic factors. With alternating layers of $AO_3$ and B units along the [111] direction, there are no oxygen in the Mn displacement plane. Partial covalent bonds between empty Mn $e_g$ states and O $2p$ states are thus weaker than under (001) strain where Mn is displaced directly towards an oxygen ion. We highlight the important electronic structure of $SrMnO_3$ for stabilizing polar Goldstone-like modes by comparing with (111) tensile strained $BaTiO_3$, where an out-of-plane polar component is predicted [42], precluding a rotationally invariant polar mode. However, when this out-of-plane polar mode remains stable, the coupling of the two in-plane polar modes in $BaTiO_3$ yields the same rotationally invariant in-plane polar mode (Fig. S4) as for $SrMnO_3$. This implies that only in the absence of an unstable out-of-plane polar mode can an in-plane polar Goldstone-like mode be realized in (111) strained perovskites. While $Ti^{4+}$ is a $d^0$ cation, the $t_{2g}$ electrons of the $d^3$ $Mn^{4+}$ cation prevents out-of-plane polar displacements of Mn, and concomitantly an unstable out-of-plane polar mode. While out-of-plane polar modes in $PbSr_2Ti_2O_7$ [14] are stabilized by confining the $6s^2$ lone pair $Pb^{2+}$ cations into single layers between non-polar rock salt layers, out-of-plane modes in (111) strained $SrMnO_3$ are stabilized by $Mn^{4+}$ with $d^3$ configuration inherent to the material.

**IV. PHASE DIAGRAM**

The equilibrium structures under (111) strain are summarized in the phase diagram presented in Fig. 4(a) along with rotational distortions projected onto pseudocubic axes, and with the magnetic ground state shown in Fig. 4(b). G-type antiferromagnetism (G-AFM) is stabilized under compressive strain, while tensile strain progressively favours ferromagnetic order, which becomes stable at ~4.5 %. The structural differences between G-AFM and FM SrMnO$_3$ are subtle and quantitative, we thus present the structural parameters calculated with G-AFM throughout the whole strain range for consistency. Condensation of the $\gamma'$ mode results in a stable $a^-a^-a^-$ $R\bar{3}c$ structure which prevails up to 4% tensile strain, with octahedral rotations increasing with strain (Fig. 4(a)). Above 4% strain a polar $Cc$ structure is stabilized, with a polarization of 36.9 $\mu C/cm^2$ at 6% strain.

Condensation of the rotational $\alpha'$ mode under compressive strain results in a monoclinic $a^-a^-c^0$ $C2/m$ structure, while condensation of $\beta'$ gives a monoclinic $a^-a^-c^-$ $C2/c$ structure (Fig. 4(a)), which are degenerate within 1 meV/f.u. The symmetry equivalent $\langle 11\bar{2}\rangle$ and $\langle 1\bar{1}0\rangle$ axes are each separated by 60 degrees, hence there are six symmetry equivalent structures with space groups $C2/c$ and $C2/m$, respectively. All structures with a combination of in-plane $\langle 11\bar{2}\rangle$ and $\langle 1\bar{1}0\rangle$ rotation axes are identified as $a^-b^-c^-$ $P\bar{1}$. The pseudocubic rotational amplitudes $\alpha, \beta$ and $\gamma$ are related for a continuous set of energy-minimizing structures by to the following equations (Fig. 4(d)): $= -\frac{1}{2}(\sqrt{3}\sin(\theta) + \cos(\theta))$, $\beta = \frac{1}{2}(\sqrt{3}\sin(\theta) - \cos(\theta))$ and $\gamma = \cos\theta$), where $\theta$ is, the shift in in-plane rotational axis as defined in Fig. 4(c).

**V. DISCUSSION AND CONCLUSION**

Having established the structural, electronic and chemical factors responsible for stabilizing Goldstone-like modes in (111) strained SrMnO$_3$, we turn to the exotic physics of having such an emergent symmetry in perovskite oxides. Goldstone mode excitations around the brim of the Mexican hat (the azimuthal direction) imply a continuous change of phase, in this case corresponding to a degenerate rotation or polarization axis. In contrast to this, excitations up the brim, corresponding to a change in mode amplitude, are called Higgs modes [44,45]. Since these modes have a finite excitation gap -- depending on the curvature of the energy potential -- the Higgs modes are massive. From our detailed analysis of the role of strain on the energy landscape, we show that the shape of the Mexican hat potential can be tuned with epitaxial strain. Therefore, in our case, the mass of a Higgs can be tuned in (111) strained $SrMnO_3$ by the choice of substrate.

The SSB in these strain-induced phase transitions is expected to result in one-dimensional topological vortices described by the Kibble-Zurek mechanism [10,11]. The first homotopy group of the order parameter space *U(1)* is non-trivial, predicting the formation of vortices in $SrMnO_3$, This is analogous to those formed in superfluid He [11], liquid crystals [5], superconductors [46,47] and hexagonal manganites [11,12,18–20], and can be understood in terms of the Kibble-Zurek scenario for topological defect formation in phase transitions.

In contrast with liquid crystals, liquid $^4$He or superconductors, the Goldstone modes, Higgs modes and vortices can be engineered by choice of substrate in epitaxially strained $SrMnO_3$. Substrates should be (111)-oriented cubic or rhombohedral to avoid non-quadratic biaxial strain which could break the symmetry of the Mexican hat potentials. A short discussion of possible substrates and the stability of $SrMnO_3$ towards oxygen vacancy formation is provided in the Supplementary Information.

In summary, we have mapped out the (111) epitaxial phase diagram for $SrMnO_3$ and established the main strain mediating mechanisms. Compressive strain gives rise to non-polar Goldstone-like modes, while large tensile strain induces a continuous polar ground state and polar Goldstone-like modes. $SrMnO_3$ is a potential condensed matter system for studying the proliferation of topological vortices in the solid state, and for tuning the amplitudes of Goldstone and Higgs modes by epitaxial strain.

**ACKNOWLEDGEMENTS**

The work is supported by NTNU and the Research Council of Norway, Grants 231430 and 231290 (A.M., M.M., T.G., T.T. and S.M.S.). S.M.G. is supported by the Director, Office of Science, Office of Basic Energy Sciences, Materials Sciences and Engineering Division, of the U.S. Department of Energy under Contract No. DE-AC02-05-CH11231 and the Swiss National Science Foundation Early Postdoctoral Mobility Program. Computational resources were provided by UNINETT Sigma2 through the projects NN9264K and NN9301K. The authors thank Quintin Meier for valuable discussions.

**COMPUTATIONAL DETAILS**

Calculations were performed using the VASP [48] code with the PBEsol [49] functional and a Hubbard U correction of 3eV applied to the 3d orbitals [50]. Plane waves were expanded up to a kinetic energy cutoff of 650 eV and sampled with a Γ-centered 6x6x2 mesh for hexagonal 30 atom cells and a Γ-centered 4x4x4 mesh for pseudocubic 40 atom cells. PAW potentials [51] treated Sr ($4s^2 4p^6 5s^2$), Mn ($3p^6 4s^2 3d^5$) and O ($s^2 p^4$) states as explicit valence electrons. Atomic positions were relaxed until forces on all ions were below 1 meV/Å. Phonon calculations were carried out on high symmetry ($R\bar{3}m$ structures and $P4/mmm$ structures for (111) and (001) strain respectively) 40 atom pseudocubic structures, obtained by rotation of relaxed hexagonal cells for each strain. G-type antiferromagnetic ordering was enforced throughout the phonon calculations. The software Phonopy [52] was used for the phonon calculations, and FINDSYM [53] was used to determine the space groups with a tolerance on the atomic positions equal to $10^{-3}$ Å. The Berry-phase formalism was used to evaluate the electronic contribution to the ferroelectric polarization [54,55].

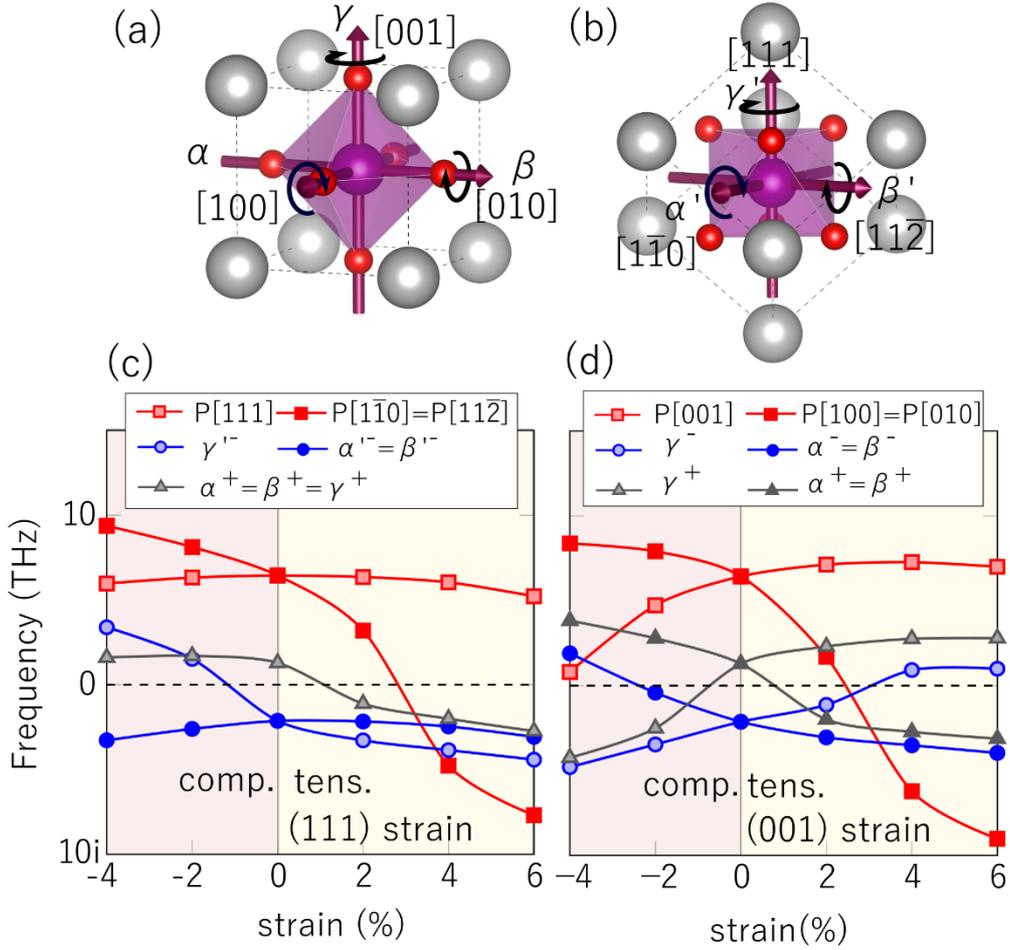

FIG. 1. Evolution of structural instabilities in $R\bar{3}m$ and $P4/mmm$ SrMnO$_3$ under (111) and (001) epitaxial strain, respectively. (a) Visualization of the pseudocubic [100], [010] and [001] axes. Octahedral rotations about the respective axes are denoted $\alpha, \beta$ and $\gamma$ (b) Visualization of the pseudocubic [1$\bar{1}$0], [11$\bar{2}$] and [111] axes, where the two former lie in the (111) plane. Octahedral rotations about the [1$\bar{1}$0], [11$\bar{2}$] and [111] axes are labelled $\alpha', \beta'$ and $\gamma'$, respectively. (c-d) Evolution of rotational and polar modes with (111) (c) and (001) (d) epitaxial strain. Polar modes are shown on red, out-of-phase rotations (⁻) are blue and in-phase rotations (⁺) are gray.

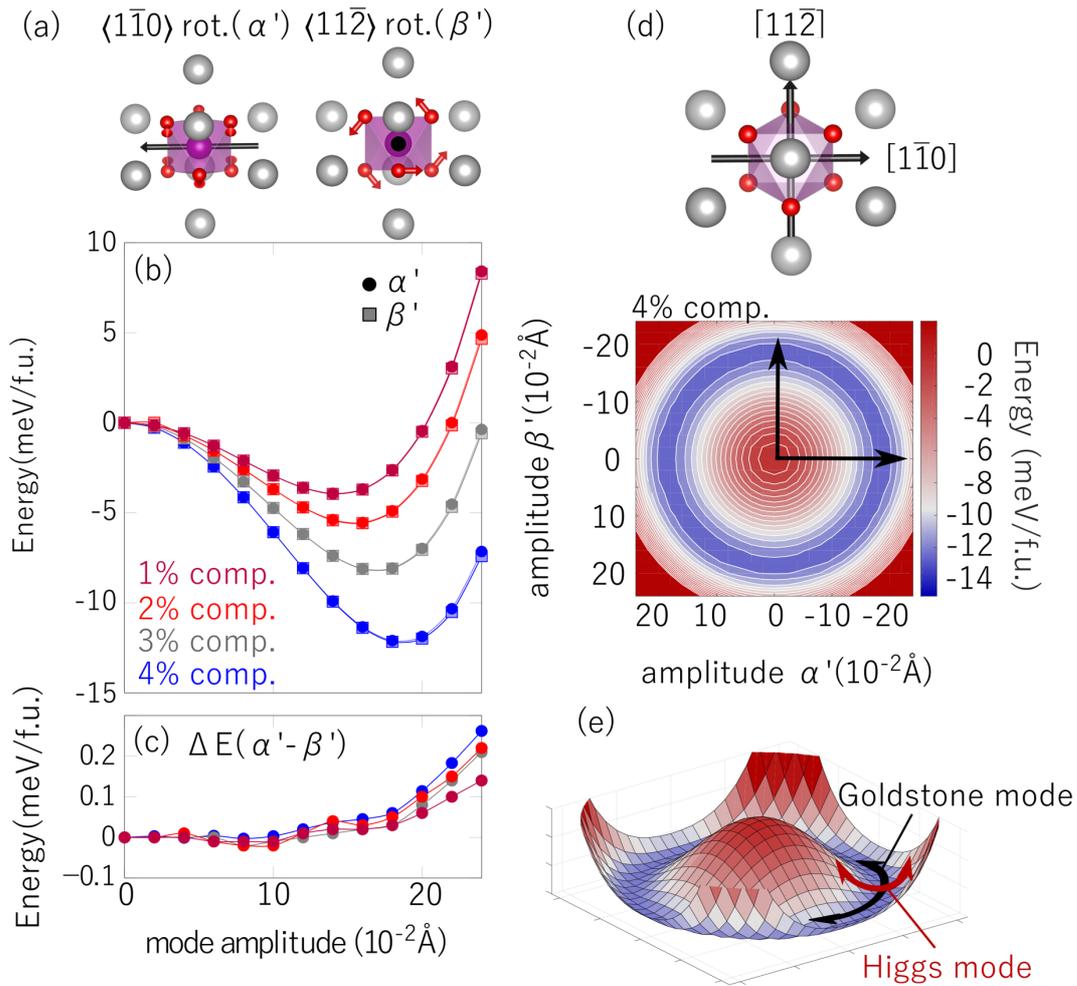

FIG. 2. Energy landscape of rotational phonon modes in SrMnO$_3$ under compressive (111) strain. (a) Atomic displacements corresponding to the two in-plane rotational modes, $\alpha'$ and $\beta'$. (b) Energy landscape $\alpha'$ and $\beta'$ modes distorted from the high symmetry R$\bar{3}$m phase, shown for compressive strain values between 1% and 4%. (c) Energy difference between $\alpha'$ and $\beta'$ modes showing an increased preference for the $\beta'$ mode with increasing amplitude. (d) Coupling of the two in-plane rotational modes at 4% compressive strain, resulting in a rotationally invariant energy landscape: a Mexican hat potential. Directions in the Mexican hat potential directly translate to in-plane rotational axes, as illustrated in the upper panel. The arrows indicate orthogonal in-plane rotational axes corresponding to the [11$\bar{2}$] and [1$\bar{1}$0] axes (e) Illustration of a Goldstone mode in which there can be massless excitations to new distinct phases along the brim of the Mexican hat.

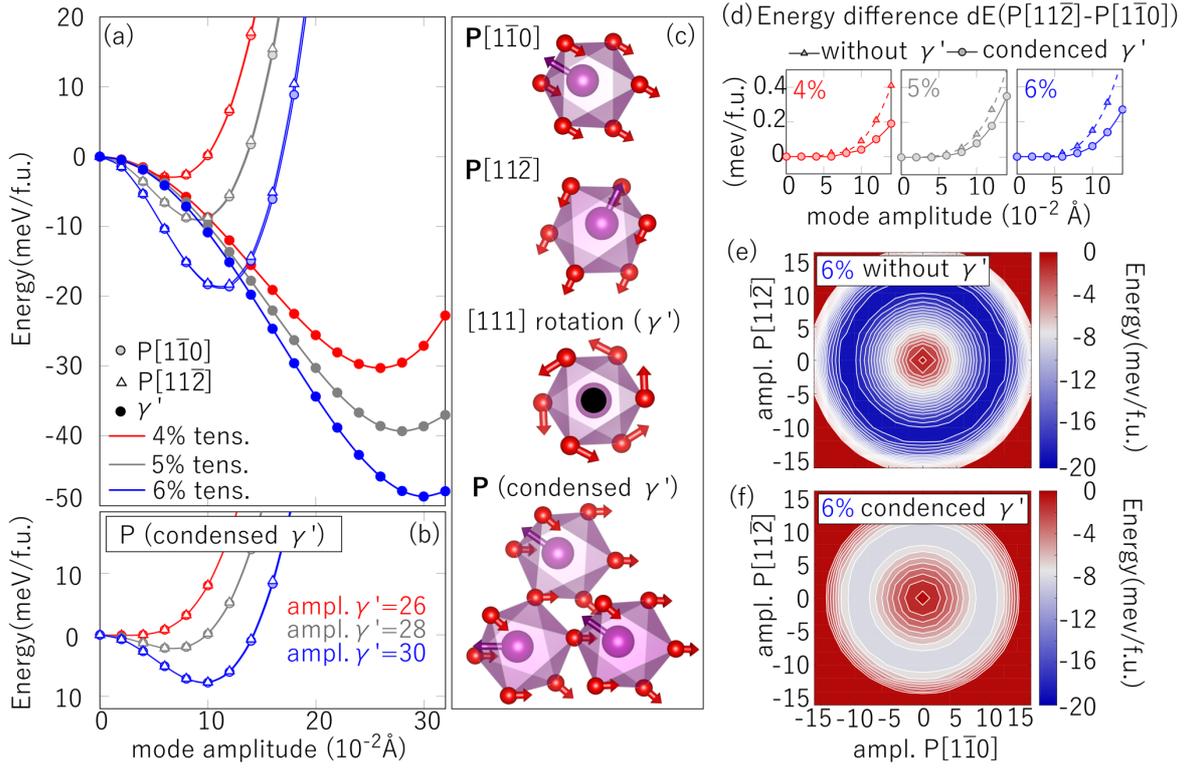

FIG. 3. Energy landscape of competing rotational and polar phonon modes in SrMnO$_3$ under tensile (111) strain. (a) Energy landscape of the rotational γ′ and the two orthogonal polar modes P[1$\bar{1}$0] and P[11$\bar{2}$], distorted from the high symmetry R$\bar{3}$m phase, for tensile strains ranging from 4% to 6%. (b) The energy landscape of polar modes when the energy minimizing rotational amplitude for each strain value is condensed. Condensed rotational amplitudes for the different strain values are indicated (red (4%), gray (5%), blue (6%) (c) Atomic displacements corresponding to the γ′ mode, the polar P[11$\bar{2}$] and P[1$\bar{1}$0] (three upper panels) and resulting polarization behavior when the γ′ mode is condensed (lower panel), in which atoms are no longer displaced along a single axis (d) The energy difference for increasing mode amplitude between the polar direction P||[11$\bar{2}$] and P||[1$\bar{1}$0], indicating an increased preference for polarization along the [1$\bar{1}$0]. Dashed lines show the energy difference in the absence of rotations, whereas solid lines show the resulting reduced energy difference when the γ′ mode is condensed. (e-f) Energy landscape upon coupling the in-plane polar modes at 6% tensile strain resulting in a Mexican hat potential energy surface with rotationally invariant ground state both in the absence of rotations (e) and in the presence of rotations frozen in the lowest energy amplitude (f).

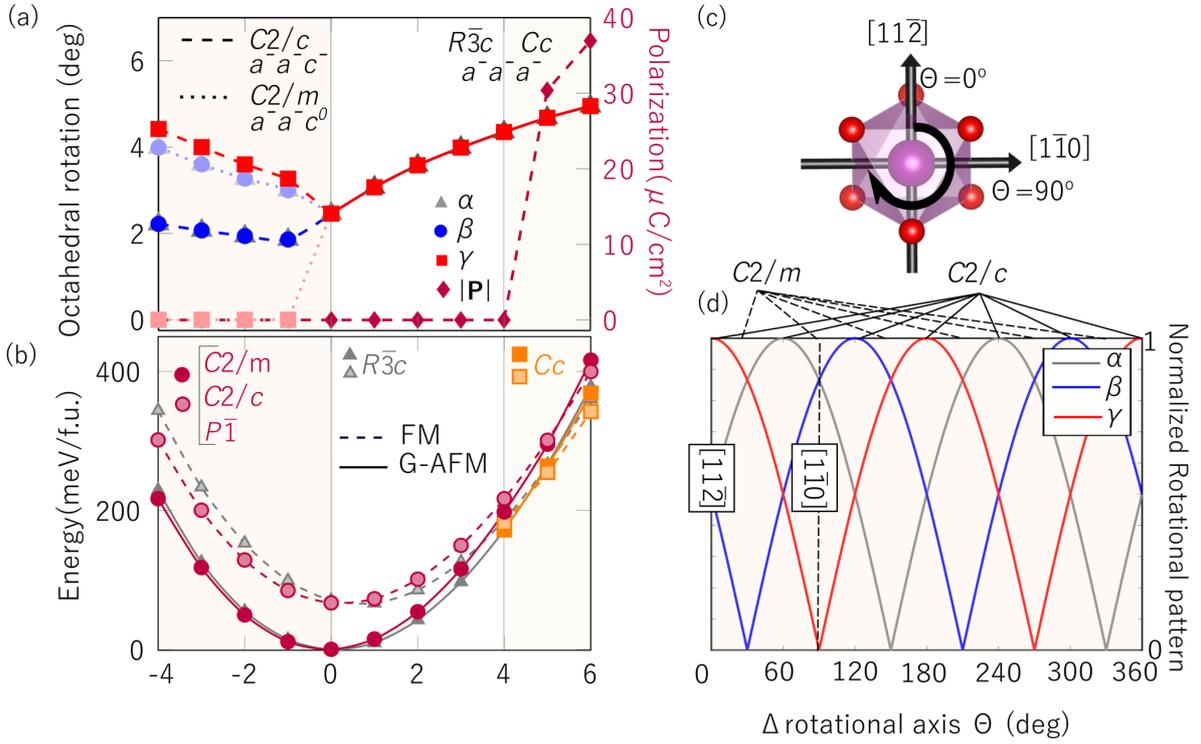

FIG. 4. Structural and magnetic phases in SrMnO$_3$ under (111) epitaxial strain. (a) Pseudocubic rotational angles and polarization as function of strain. Under compressive strain the *C*2/*c* phase (rotational angles given by dashed lines) and the *C*2/*m* (rotational angles given by dotted lines) are degenerate. (b) Total energy as function of strain for different structural phases, comparing the stability of G-AFM ordering with FM ordering. Energies are given per formula unit with the *R*$\bar{3}$*c* unstrained G-AFM structure as the reference state. A magnetic crossover from G-AFM to FM is found at ~4.5 % tensile (111) strain (c) Definition of the shift θ in in-plane rotational axis. θ = 0 corresponds to the rotational axis [11$\bar{2}$] θ = 90 corresponds to the [1$\bar{1}$0] direction. (d) The analytically derived relation between the rotational angles α, β and γ upon a shift (θ) of the in-plane rotational axis. Total rotational angles are normalized.

# Supplementary information

## Goldstone-like phonon modes in a (111)-strained perovskite

A. Marthinsen[1], S. M. Griffin[2,3], M. Moreau[4], T. Grande[1], T. Tybell[4], and S. M. Selbach[1,*]

[1] Department of Materials Science and Engineering, NTNU Norwegian University of Science and Technology, 7491 Trondheim, Norway
[2] Molecular Foundry, Lawrence Berkeley National Laboratory, Berkeley, California, USA.
[3] Department of Physics, UC Berkeley, Berkeley, California, USA.
[4] Department of Electronic Systems, NTNU Norwegian University of Science and Technology, 7491 Trondheim, Norway

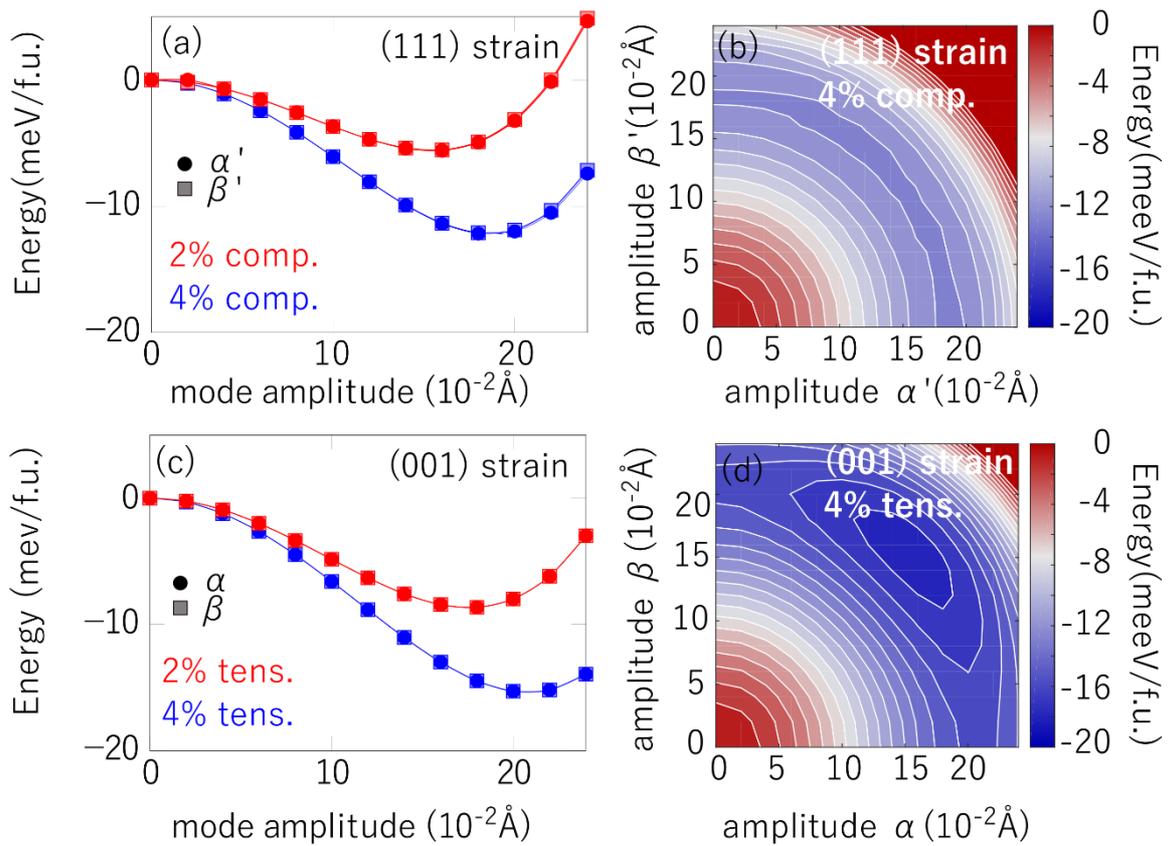

FIG. S1. (a) Energy curves for freezing in in-plane rotational modes under compressive (111) strain. (b) Coupling of the two in-plane rotational modes resulting in a Goldstone-like non-polar energy landscape. (c) Energy curves of in-plane rotational modes for under tensile (001) strain. (d) Coupling of the two in-plane rotational mode at 4% tensile strain resulting in a distinct energy minimum with [110] rotational axis.

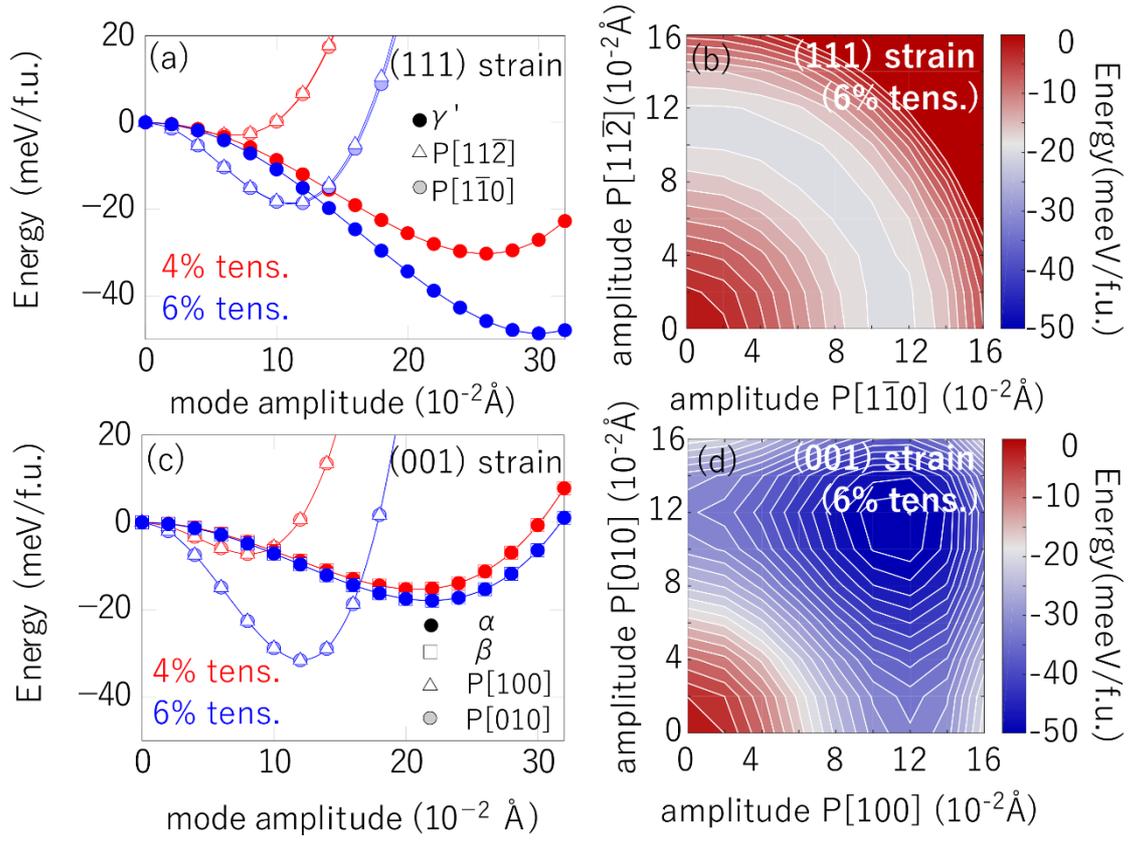

FIG. S2. (a) Energy curves resulting from distortion by polar and rotational modes from high symmetry $R\bar{3}m$ phases under (111) strain. (b) The coupling of the two in-plane polar distortions at 6% strain yields a Goldstone-like energy landscape. (c) Energy curves resulting from distortion by polar and rotational modes from high symmetry $P4/mmm$ phases under (001) strain. (d) The coupling of the two in-plane polar distortions at 6% strain yields a distinct minimum energy polarization direction corresponding to [110].

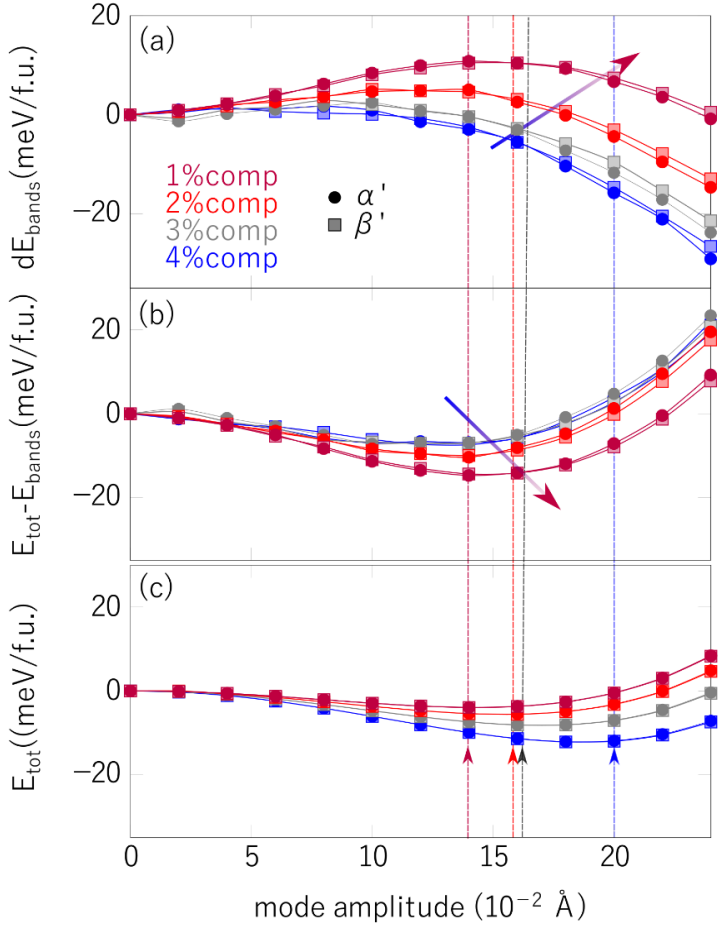

FIG. S3. Energy contributions (a-b) to the total energy (c) of the rotational $\alpha'$ and $\beta'$ modes with increasing compressive strain ranging from 1% to 4%. (a) Band energies accounting for the electronic contribution to the total energy. The band contributions go through a maximum before monotonically decreasing for higher amplitudes. Both the maximum and amplitude at which this peak occurs are reduced with increasing strain. (b) The electrostatic energy contribution calculated by the difference between total energy and band energy. The electrostatic energy goes through a minimum before increasing, and is thus the limiting contribution to the amplitudes of the total energy minima. For increasing compressive strain, the electrostatic gain from increased rotation is reduced. (c) The total energy. Amplitudes of the energy minima are indicated with dotted lines. The amplitudes and energy gain of the total energy minima show the opposite trend with strain as the electrostatic contribution indicating that gain in the band energy account for the increased mode amplitude with increasing compressive strain.

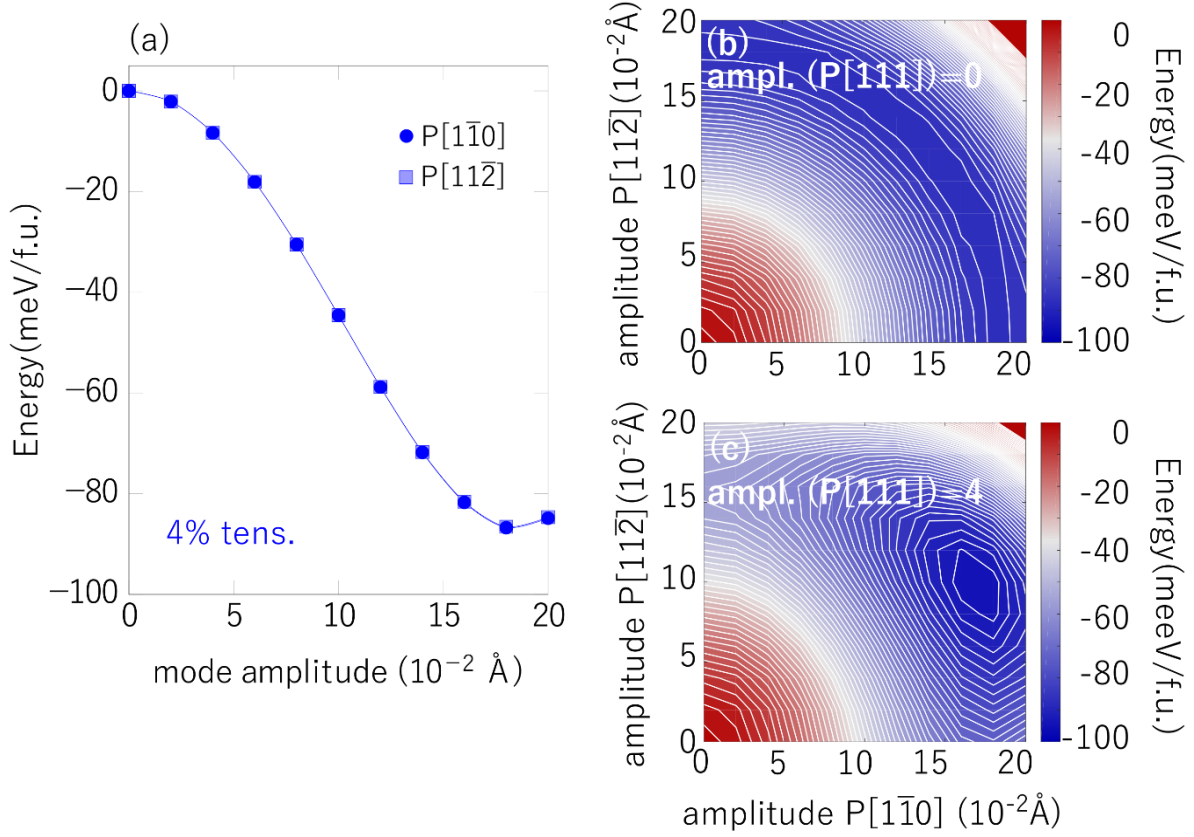

FIG. S4. Coupling of the two orthogonal in-plane polar modes in BaTiO$_3$ 4% tensile (111) strain. (a) Energy curves resulting from distortion by polar modes P[1$\bar{1}$0] and P[11$\bar{2}$] from high symmetry $R\bar{3}m$ phases under (111) strain. (b) Coupling of modes P[1$\bar{1}$0] and P[11$\bar{2}$] distorted from high symmetry $R\bar{3}m$ (no out-of-plane polarization) yielding a rotational invariant Mexican hat shaped energy potential, in analogy with the polar modes in SrMnO$_3$. (c) Coupling of modes P[1$\bar{1}$0] and P[11$\bar{2}$] when out-of-plane [111] polarization of amplitude 4*10$^{-2}$ Å is frozen in. The condensation of the [111] polar mode, breaks the hexagonal symmetry – resulting in trigonal symmetry and a distinct energy minima along the [2$\bar{1}\bar{1}$] direction.

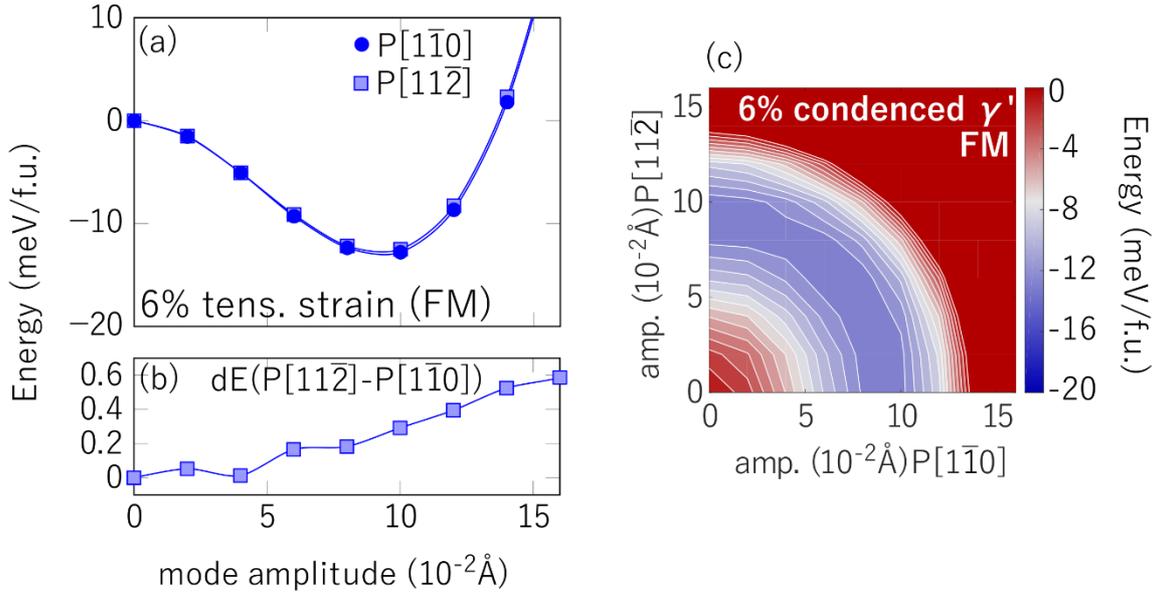

FIG. S5. Energy curves resulting from distortion by polar modes from a relaxed $R\bar{3}c$ phase with FM ordering under at 6% tensile (111) strain, showing the same qualitative behavior as with G-AFM ordering. (b) The coupling of the two in-plane polar distortions at 6% strain yielding a Goldstone-like energy landscape. (c) Energy difference between polar directions $P[11\bar{2}]$ and $P[1\bar{1}0]$ as function of mode amplitude, showing preference for the $P[1\bar{1}0]$ direction, but only an energy difference of ~0.3meV at the energy minimizing amplitude.

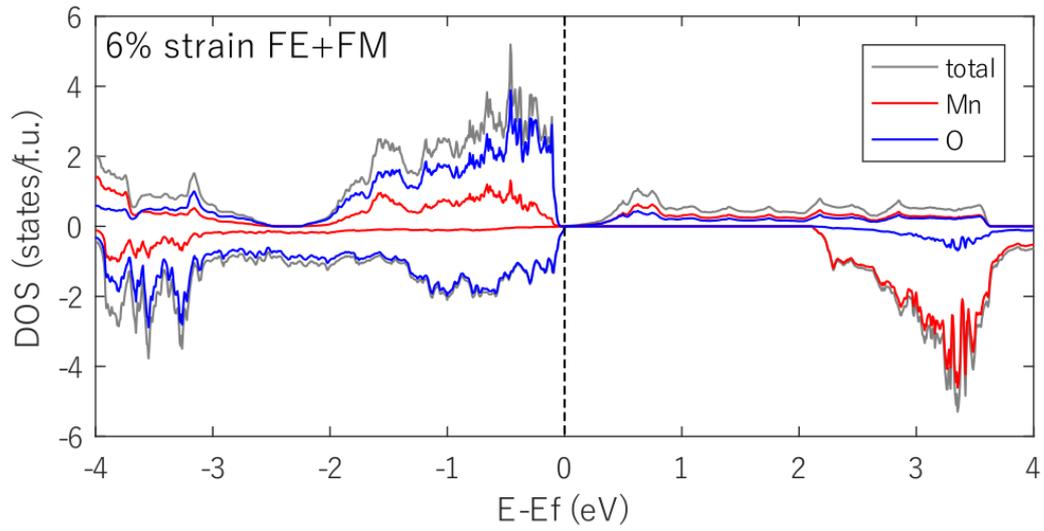

FIG. S6. Density of states of 6% strained SrMnO$_3$ with FM ordering and polar distortion yielding a zero-gap insulating state. The vertical dotted line indicates the fermi level.

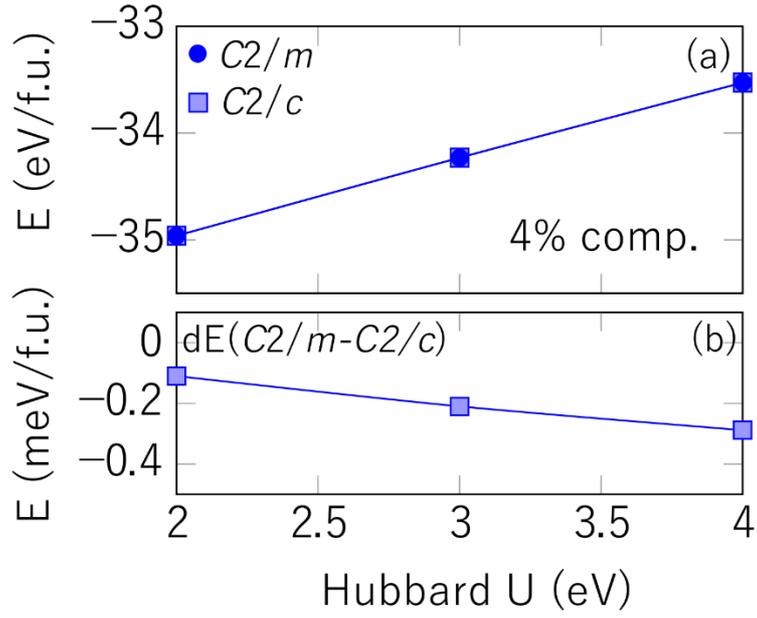

FIG. S7. Effect of a Hubbard U potential on Mn *d* states on (a) the total energy and (b) the energy difference between *C*2/*c* and *C*2/*m* phases at 4% compressive strain. Whereas the total energy is significantly increased when the correlation is increased, the energy difference between the *C*2/*c* and *C*2/*m* phases are only slightly altered and well below 1meV.

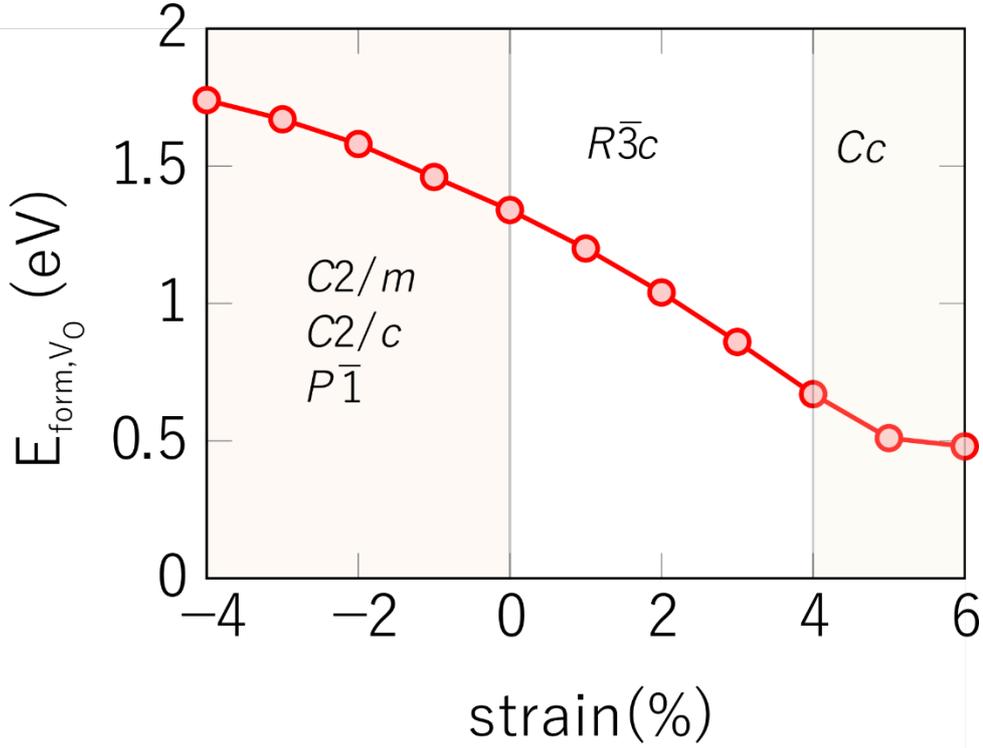

FIG. S8. Oxygen vacancy formation ($E_{form,V_O}$), given by $E_{form,V_O} = E_{tot,V_O} - E_{Stoich} + \mu_O$, as function of epitaxial (111) strain. The oxygen chemical potential $\mu_O = -5eV$, corresponds to typical growth conditions under air. The formation energy is calculated by removing one oxygen from a 40 atom supercell, yielding a stoichiometry of SrMnO$_{2.875}$. Coherent (111) strain does not give rise to inequivalent oxygen sites. Inequivalent oxygen sites only arise as a result of structural distortions in the monoclinic phases under compressive strain and in the polar phase, but since the differences are small (differing in energy by an order of magnitude of ~0.01eV), only the average Vo formation is shown. In the presence of an oxygen vacancy, FM ordering is more energetically stable than G-AFM ordering, within a 40 atom supercell for all strain values investigated.

**Supplementary note 1: Derivation of pseudocubic rotational angles $\alpha, \beta$ and $\gamma$ under compressive strain**

The octahedral rotations $\alpha, \beta$ and $\gamma$ about pseudocubic axes [100], [010] and [001], respectively, are found to be proportional to the elements in vector [x,y,z] parallel to the in-plane rotational axis

$$[\alpha, \beta, \gamma] = n[x, y, z] \quad (1)$$

Given that the amplitude of distortion around the brim of the Mexican hat is constant, this further implies that

$$\alpha^2 + \beta^2 + \gamma^2 = A^2 \quad (2)$$

where *A* is the amplitude of the total rotational distortion. Furthermore, given that the in-plane rotational vectors [x,y,z] are all perpendicular to the [111] direction, this gives the constraint

$$\alpha + \beta + \gamma = 0 \quad (3)$$

Constraint (2) spans a sphere in the phase space spanned by $\alpha, \beta$ and $\gamma$, whereas constraint (3) spans a plane. We combine (2) and (3) to get:

$$\beta^2 + \gamma^2 + (\beta+\gamma)^2 = \left(\beta+\frac{\gamma}{2}\right)^2 + \frac{3}{4}\gamma^2 = A^2 \Rightarrow \frac{\left(\beta+\frac{\gamma}{2}\right)^2}{A^2} + \frac{\gamma^2}{\frac{4}{3}A^2} = 1 \quad (4)$$

which spans an ellipse. We can parameterize the ellipse according to:

$$\beta + \frac{\gamma}{2} = A * \sin(\theta), \theta = [0, 2\pi] \quad (5a)$$

$$\gamma = \frac{2A}{\sqrt{3}} * \cos(\theta), \theta = [0, 2\pi], \quad (5b)$$

Where the variable $\theta$ corresponds to a rotation of the in-plane rotational axis, and the $[11\bar{2}]$ direction is (arbitrarily) chosen to be $\theta = 0$. Combining (3), (5a) and 5b) yields

$$\alpha = -A(\sin\theta + \frac{1}{\sqrt{3}}\cos\theta) \quad (6a)$$

$$\beta = A(\sin(\theta) - \frac{1}{\sqrt{3}}\cos\theta) \quad (6b)$$

$$\gamma = \frac{2A}{\sqrt{3}} * \cos(\theta) \quad (6c)$$

Normalizing these equations yields:

$$\alpha = -\frac{1}{2}(\sqrt{3}\sin\theta + \cos\theta) \quad (7a)$$

$$\beta = \frac{1}{2}(\sqrt{3}\sin(\theta) - \cos\theta) \quad (7b)$$

$$\gamma = \cos(\theta) \quad (7c)$$

**Supplementary note 2: Substrates**

We list possible substrates for growth of SrMnO3 ($a_0$=3.800Å) in Table S1, based on the lattice parameter of the substrate, that would produce a range of different epitaxial strain values in SrMnO$_3$. We note that substrates with rhombohedral space group are more promising than orthorhombic substrates, as the latter may remove the degeneracy along the brim of the Mexican hat potential. Furthermore, only a few of these materials are reported[1] to be commercially available as substrates, including YAlO$_3$ , NdAlO$_3$ and KTaO$_3$.

**Table S1 Potential substrate materials**

| Substrate | Space group (Ground state) | Lattice parameter | Strain (%) | |
|---|---|---|---|---|
| MgTiO$_3$ | (R3-H)Ilmenite[2] | 3,57[2] | -5,95 | |
| **LiNbO$_3$** | *R3c*[3] | **3,64**[3] | **-4,21** | |
| **LiTaO$_3$** | *R3c*[4] | **3,64**[4] | **-4,09** | |
| GdAlO$_3$ | *Pnma*[5] | 3,71[5] | -2,37 | |
| YAlO$_3$ | *Pnma*[6] | 3,72[1] | -2,11 | |
| SmCoO$_3$ | *Pnma*[7] | 3,75[7] | -1,32 | |
| **NdAlO$_3$** | *R-3c*[8] | **3,76**[1] | **-0,94** | |
| **BiFeO$_3$** | *R3c*[9] | **3,94**[9] | **3,79** | |
| SrMoO$_3$ | *Imma*[10] | 3,98[10] | 4,61 | I4/mcm(266K)-Imma(125K)) |
| **KTaO$_3$** | *Pm-3m*[11] | **3,99**[1] | **4,95** | Volatile at elevated T, therefore avoided in some substate applications |
| **KNbO$_3$** | *R3m*[12] | **4,01**[12] | **5,45** | |
| **BaTiO$_3$** | *R3m*[13] | **4,01**[13] | **5,58** | |
| SrSnO$_3$ | *Pnma*[14] | 4,03[14] | 6,16 | |
| SrHfO$_3$ | *Pnma*[15] | 4,07[15] | 7,08 | |
| SrZrO$_3$ | *Pnma*[16] | 4,10[16] | 7,92 | |
| BaSnO$_3$ | *Pm-3m*[17] | 4,12[17] | 8,32 | |
| BaZrO$_3$ | *Pm-3m*[18] | 4,19[18] | 10,26 | |
| LaLuO3–LaScO3 | *Pnma*[19] | 4,05-4,18[19] | 6,58-10,00[19] | Solid solution |

**Supplementary note 3: Oxygen vacancy formation**

The oxygen vacancy formation energy as function of epitaxial (111) strain (Figure S8) follow the same trend as for (001) strain[20], being lowered with tensile strain as a reverse and analogue effect of chemical expansion[21], whereas under compressive strain the formation energy increases. As also previously reported for (001) strain[20], a polar distortion competes with the formation of oxygen vacancies, resulting in a decrease of the slope at the phase boundary. Importantly, the oxygen vacancy formation energy stays positive and above ~0.5eV in the whole strain range considered in this study, indicating that formation of oxygen vacancies is *not* enthalpy stabilized. Re-oxidation at low temperatures of the thin film after growth is thereby a possibility to achieve stoichiometric thin films.